\documentclass[a4paper,11pt]{article}
\pdfoutput=1 % if your are submitting a pdflatex (i.e. if you have
\usepackage{jcappub} % for details on the use of the package, please
\usepackage[T1]{fontenc} % if needed

\title{Dependence of Galaxy Stellar Properties on the Primordial Spin Factor}

\author[a,b]{Jun-Sung Moon}
\author[a,1]{, Jounghun Lee \note{Corresponding author.}}

\affiliation[a]{Department of Physics and Astronomy, Seoul National University, \\
Kwanak-ro 1, Kwanak-gu, Seoul 08826, Republic of Korea}
\affiliation[b]{Research Institute of Basic Sciences, Seoul National University,  \\
Kwanak-ro 1, Kwanak-gu, Seoul 08826, Republic of Korea}

\emailAdd{jsmoon.astro@gmail.com}
\emailAdd{cosmos.hun@gmail.com}

\abstract{We present a numerical discovery that the observable stellar properties of present galaxies retain significant dependences on the primordial density and tidal 
fields. Analyzing the galaxy catalogs from the TNG300-1 simulations, we first compute the primordial spin factor, $\tau$, defined as the mean degree of misalignments 
between the principal axes of the initial density and potential hessian tensors at the protogalactic sites. Then, we explore in the framework of Shannon's information theory if and 
how strongly each of six stellar properties of the present galaxies, namely two stellar sizes ($R_{90\star}$ and $R_{50\star}$), ages, specific star formation rates, optical colors 
and metallicities, share mutual information with $\tau$, measured at $z=127$. 
Deliberately controlling the TNG galaxy samples to have no differences in the mass, environmental density and shear distributions, we single out net 
effects of $\tau$ on each of the galaxy stellar properties. In the higher stellar mass range of $M_{\star}/(h^{-1}\,M_{\odot})\ge 10^{10}$, significant amounts of 
mutual information with $\tau$ are exhibited by all of the six stellar properties, while in the lower range of $M_{\star}/(h^{-1}\,M_{\odot})< 10^{10}$ only four of the six properties 
except for the specific star formation rates and colors yield significant signals of $\tau$-dependence. Examining how the mean values of the six stellar properties vary with $\tau$, 
we also show that the galaxies originated from the protogalactic sites with higher $\tau$ values tend to have larger sizes, later formation epochs, higher specific star formation rates, 
bluer colors and lower metallicities. It is also discovered that the galaxy stellar sizes, which turn out to be most robustly dependent on $\tau$ regardless of $M_{\star}$, follow 
a {\it bimodal} Gamma distribution, the physical implication of which is discussed.} 
\begin{document} 
\maketitle
\flushbottom

\section{Introduction}\label{sec:intro}

It was conventionally believed that the galaxy stellar properties have little connection with initial conditions of the universe as they should be mainly established through 
essentially stochastic processes in the subsequent evolution~\cite{sei-etal79,sta-etal83,CT19,tac-etal20,wan-etal22,PF23}. 
This conventional notion implied that it would be almost impossible to probe the early universe physics with observable galaxy properties. 
Several recent numerical studies based on high-resolution hydrodynamic simulations, however, have indicated that this notion may be too pessimistic, demonstrating 
that certain observable galaxy properties associated with the total angular momenta of host dark matter (DM) halos in fact vary sensitively with initial 
conditions~\cite{sal-etal12,KL13,shi-etal17,yu-etal20,cad-etal22,per-etal22,ML23,she-etal24}. 

These numerical indications are in line with the observational detection of ref.~\cite{dis-etal08} that the mutual correlations among the key stellar properties of local 
spiral galaxies like the stellar masses, sizes, ages, metallicities, and colors appeared to be much simpler than expected, the results of which were supported 
by several follow-up works~\cite{gar-etal09,cha-etal12,sha-etal23}.  It was originally suggested by ref~\cite{dis-etal08} that the simple correlation structure among the key stellar 
properties should be interpreted as a counter-evidence against the standard theory of hierarchical structure formation scenario in which individual galaxies undergo haphazardly 
different physical processes~\cite{GT_review}. Later, ref.~\cite{ML24a}  explained that the observed simple correlation structure of galaxy stellar properties must reflect how much 
impact the halo angular momenta have on the galaxy intrinsic properties rather than implying any inconsistency with the hierarchical structure formation scenario.

It was revealed by several numerical and observational works that the galaxy properties like the formation epochs, sizes, surface brightnesses and morphologies 
are dependent not only on the total masses of DM halos but also on their total angular momenta~\cite{ber-etal08,KL13,tek-etal15,rod-etal22,per-etal22,du-etal22,rom-etal23,ML24a}. Besides,  multiple observational studies also proved that the galaxy angular momenta still retain significant amounts of memory for the initial conditions of the early 
universe~\cite{mot-etal21,mot-etal22,ML23}, as predicted by the linear tidal torque theory~\cite{dor70,whi84,LP00,LP01}. As pointed out by~\cite{ML23}, even though the halo angular momenta develop deviations from the predictions of tidal torque theory through hierarchical merging processes which indeed have an effect of modifying 
the directions and magnitudes of halo angular momenta from the protohalo versions~\cite{vit-etal02}, 
their connections with the initial tidal fields do not severely diminish, as the spin angular momenta of merged galaxies acquire connections with the initial tidal fields 
on larger scales through the process of orbital angular momenta transfer. 

Given this existence of interdependence among the initial conditions, halo angular momenta and galaxy properties, 
ref.~\cite{ML24a} developed an efficient way to quantify the effects of initial conditions on the galaxy properties in the context of Shannon's information theory~\cite{sha48}. 
Defining a single parameter initial condition, $\tau$, as the degree of misalignments between the principal axes of the protogalaxy inertia and initial tidal tensors~\cite{ML24b}, 
to which the first order generations of protohalo angular momenta are critically subject~\cite{whi84,LP00},  they numerically evaluated the amounts of mutual information 
between $\tau$ and various intrinsic properties of galactic halos as measures of the interdependences. It was found in their study that the formation epochs, spin parameters, 
stellar to total mass ratio and kinematic morphologies indeed share significants amounts of mutual information with this single parameter initial condition. 

Nevertheless, the intrinsic properties of galactic halos considered in the work of ref.~\cite{ML24a} were not directly observable ones. Furthermore, it was not properly taken 
into account that the significant amounts of mutual information detected by~\cite{ML24a} between $\tau$ and intrinsic properties of present galactic halos could be at least 
partially contributed by the differences in the total mass and environments, which have been well known to affect the formation epochs and evolutionary tracks of galactic halos 
\citep{PG84,bos-etal03,tan-etal04,sha-etal06,LL08,lee18,SP20,PS20,nan-etal24}. 
Another difficulty in applying the methodology of ref.~\cite{ML24a} to observations comes from the fact that it is impossible to evaluate from real data the single parameter 
initial condition $\tau$, which is defined in terms of the inertia tensors of protogalactic sites.

Our task here is to answer the following vital questions. Is there a more practical and feasible way to define the single parameter initial condition? 
Are the {\it observable stellar} properties of present galaxies also significantly dependent on the single parameter initial condition? Is it possible to single out the 
$\tau$-dependence of galaxy stellar properties free from the dominant effects of total mass and environments? To conduct this task, we will take the same information 
theoretical approach as in our prior work~\cite{ML24a}. 

\section{MI between the primordial spin factor and galaxy stellar properties}

\subsection{Primordial spin factor as the single parameter initial condition}\label{sec:tau}

%%%%%%%%%%%%%%%%%%%%%%%%%%%%%%%%%%%%%%%%%%%%%%%%%%%%%%%%%%%%%%%%
\begin{figure}[tbp]
\centering % \begin{center}/\end{center} takes some additional vertical space
\includegraphics[width=0.85\textwidth=0 380 0 200]{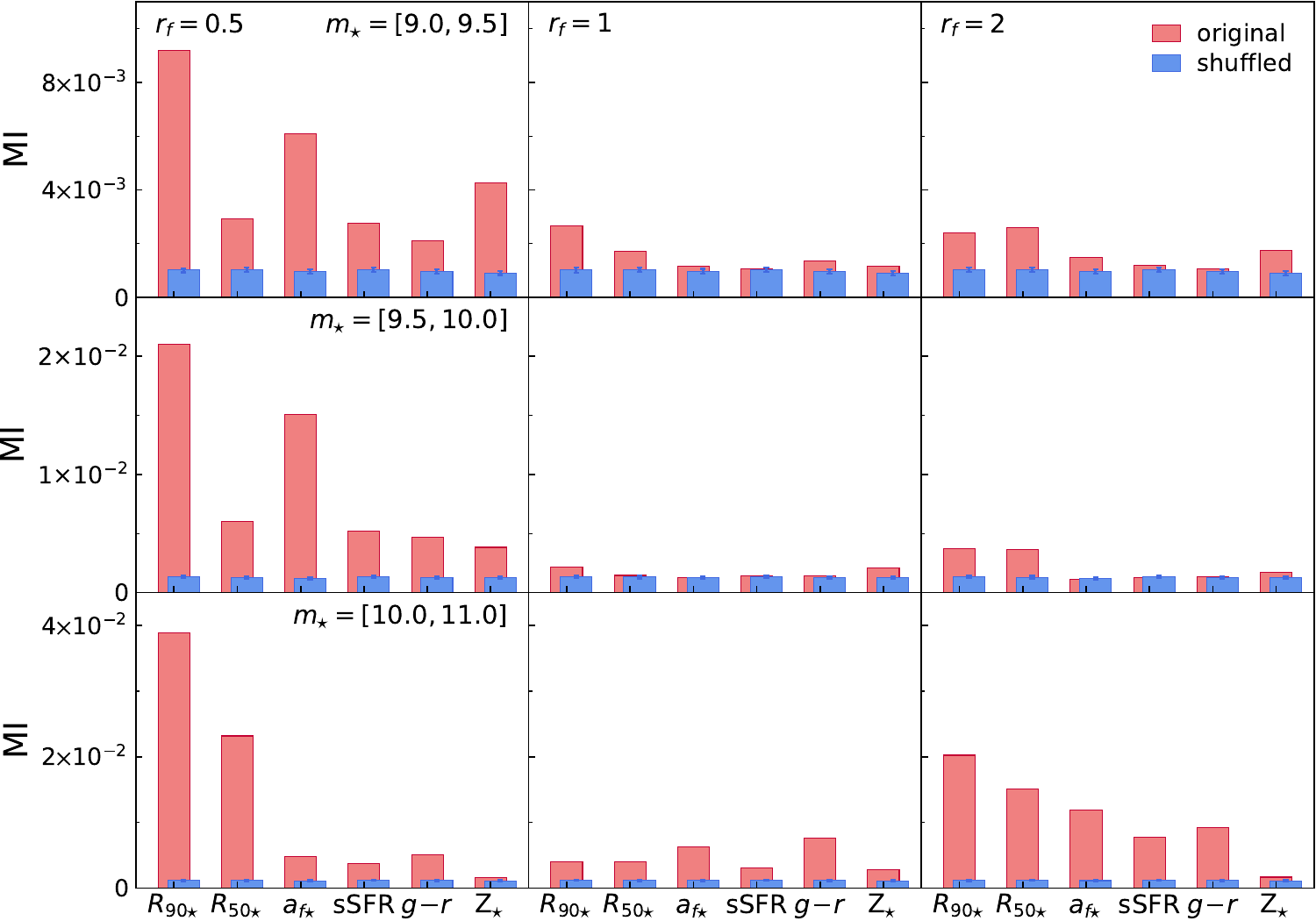}
% "\includegraphics" is very powerful; the graphicx package is already loaded
\caption{\label{fig:ori}  Mutual information (MI) values (red histogram) between six stellar properties of the TNG central galaxies at $z=0$ and the primordial 
spin factor, $\tau$, measured at $z_{i}=127$ in three different stellar mass ranges for three different cases of the Lagrangian scale, $r_{f}$ on 
which the initial density and potential hessian tensor fields are smoothed. Average MI values obtained from the $1000$ resamples with 
shuffled ($S,\ \tau$) (blue histogram) are also plotted with errors for comparison.}
\end{figure}
%%%%%%%%%%%%%%%%%%%%%%%%%%%%%%%%%%%%%%%%%%%%%%%%%%%%%%%%%%%%%%%%

To study if and how strongly the stellar properties of galactic subhalos depend on the single parameter initial condition, we utilize the particle snapshots and subhalo catalogs at 
$z=0$ from the 300 Mpc volume run (TNG300-1) of the IllustrisTNG suite of cosmological hydrodynamical simulations~\citep{tng1,tng2,tng3,tng4,tng5, tng6}, which have already 
been released to the general public\footnote{https://www.tng-project.org/data/}. 
The TNG300-1 run, performed on a period box of side length $L_{\rm tot}\equiv 205\,h^{-1}{\rm Mpc}$ for the Planck $\Lambda$CDM cosmology~\cite{planck15}, 
simulated the gravitational and hydrodynamical dynamics of $2500^{3}$ DM particles and equal number of baryonic cells with individual masses, 
$5.9$ and $1.1$, in unit of $10^{7}\,M_{\odot}$, respectively, by implementing the competent AREPO code~\cite{arepo}. 

The TNG300-1 subhalo catalogs contain the substructures of  friends-of-friends (FoF) groups identified via the SUBFIND algorithm~\cite{subfind}, providing various information 
on their intrinsic and stellar properties like their comoving positions (${\bf x}_{c}$), total masses ($M_{\rm t}$), stellar masses ($M_{\star}$), 
stellar formation epochs ($a_{f\star}$), specific star formation rate (sSFR), photometric magnitudes at eight different wavebands ($U,\ B,\ V,\ K, \ g,\ r,\ i,\ z$), stellar metallicities ($Z_{\star}$) as well as the 
comoving positions and masses of each constituent particle ($\{{\bf x}_{\alpha},\ m_{\alpha}\})$, respectively. 
Here, the stellar formation epoch, $a_{f\star}$, is defined as the scale factor at the mean age of constituent stars.
Among the central subhalos of the TNG FoF groups, only those that contain more than $100$ stellar particles are included in our main sample of the target galaxies~\cite{bet-etal07}. 
Throughout this paper, we will consider six different stellar properties of each target galaxy, $R_{90\star}$, $R_{50\star}$, $a_{f\star}$, sSFR, $g-r$ colors, and $Z_{\star}$, 
where  $R_{90\star}$ and $R_{50\star}$, denote two different stellar sizes determined as the radial distances which enclose $90\%$ and  
 $50\%$ of $M_{\star}$,  being called the stellar 90 and 50 percent-mass radii, respectively. 

For each target galaxy in the main sample, we locate the initial positions, $\{{\bf q}_{\alpha}\}$, of its constituent DM and gas particles at the initial redshift 
$z_{i}=127$, and compute their center of mass, $\bar{\bf q}\equiv M^{-1}_{\rm t}\sum_{\alpha}m_{\alpha}{\bf q}_{\alpha}$ as the protogalactic site. 
Then, we determine the primordial spin factor, $\tau(\bar{\bf q})$, at each protogalactic site by taking the following steps:  
%%%%%%%%%%%%%%%%%%%%%%%%%%%%%%%%%%%%%%%%%%%%%%%%%%%%%%%%%%%%%%%%%%%%%%%%%%%%%%%%%%%%
\begin{enumerate}
\item
Reconstruct the initial raw density contrast field, $\delta({\bf q})$, on $512^{3}$ grid points from the particle snapshot at  $z_{i}$ 
via the cloud-in-cell algorithm. 
\item
Compute the Fourier-space raw density contrast field, $\delta({\bf k})$, by using the fast Fourier transformation (FFT) code, where ${\bf k}$ is the Fourier-space 
wave vector.
\item
Compute the density hessian tensor smoothed with a Gaussian kernel on the scale of $R_{f}$ as 
$H_{ij}({\bf k})\equiv k_{i}k_{j}\delta({\bf k})\exp\left(-k^{2}R^{2}_{f}/2\right)$ where $k\equiv \vert{\bf k}\vert$.
\item
Compute the Gaussian filtered tidal tensor as $T_{ij}({\bf k})=H_{ij}({\bf k})/k^{2}$, which is equivalent to the initial potential hessian tensor.
\item
Conduct inverse Fourier transformations of the smoothed density hessian and tidal tensor fields into the real space counterparts, $H_{ij}({\bf q})$ 
and $T_{ij}({\bf q})$, respectively. 
\item
Find the orthonormal eigenvectors of $T_{ij}(\bar{\bf q})$ at each protogalactic site, and express $H_{ij}(\bar{\bf q})$ in the principal frame of $T_{ij}(\bar{\bf q})$ 
via a similarity transformation
\item
Compute the primordial spin factor, $\tau(r_{f})$, as 
%%%%%%%%%%%%%%%%%%%%%%%%%%%%%%%%%%%%%%%%%%%%%%%%%%%%%%%%%%%%%%%%%%%%%%%%%%%
\begin{equation}
\label{eqn:tau}
\tau \equiv \left(\frac{H^{2}_{12}+H^{2}_{23}+H^{2}_{31}}{H^{2}_{11}+H^{2}_{22}+H^{2}_{33}}\right)^{1/2}\, .
\end{equation}
%%%%%%%%%%%%%%%%%%%%%%%%%%%%%%%%%%%%%%%%%%%%%%%%%%%%%%%%%%%%%%%%%%%%%%%%%%
\end{enumerate}
 
It is important to note the difference between eq.~(\ref{eqn:tau}) and the original definition of $\tau$ given in our prior works~\cite{ML24a,ML24b}. 
Basically, eq.(\ref{eqn:tau}) replaces the protogalaxy inertia tensor, $(I_{ij})$, by the density hessian matrix , $(H_{ij})$, on the ground that $(H_{ij})$ was numerically found to 
approximate $(I_{ij})$ quite well as long as $R_{f}$ is comparable to the protogalaxy Lagrangian size~\cite{yu-etal20}. 
There are two main advantages of using $(H_{ij})$ instead of $(I_{ij})$ for $\tau$. First, the former can be uniquely defined unlike the latter, which depends sensitively on the 
subhalo identification algorithm, i.e., subhalo boundary. Second, it is in principle possible to statistically reconstruct the former from real observational 
data~\cite{LP00,yu-etal20,mot-etal21,mot-etal22}, while the latter is not practically measurable. 

The non-negligible value of $\tau$ is the sole initial condition for the first order generation of protogalaxy angular momenta~\cite{whi84}. From here on, this single parameter 
initial condition, $\tau$, will be called  {\it the primordial spin factor}, and its mutual information with the six stellar properties of the TNG galaxies at $z=0$ will be determined for 
three different cases of the smoothing scale, $r_{f}\equiv R_{f}/(h^{-1}{\rm Mpc})=0.5$, $1$, and $2$ in sections~\ref{sec:nocon}-\ref{sec:con}. 

%%%%%%%%%%%%%%%%%%%%%%%%%%%%%%%%%%%%%%%%%%%%%%%%%%%%%%%%%%%%%%%%
\begin{figure}[tbp]
\centering % \begin{center}/\end{center} takes some additional vertical space
\includegraphics[width=0.85\textwidth=0 380 0 200]{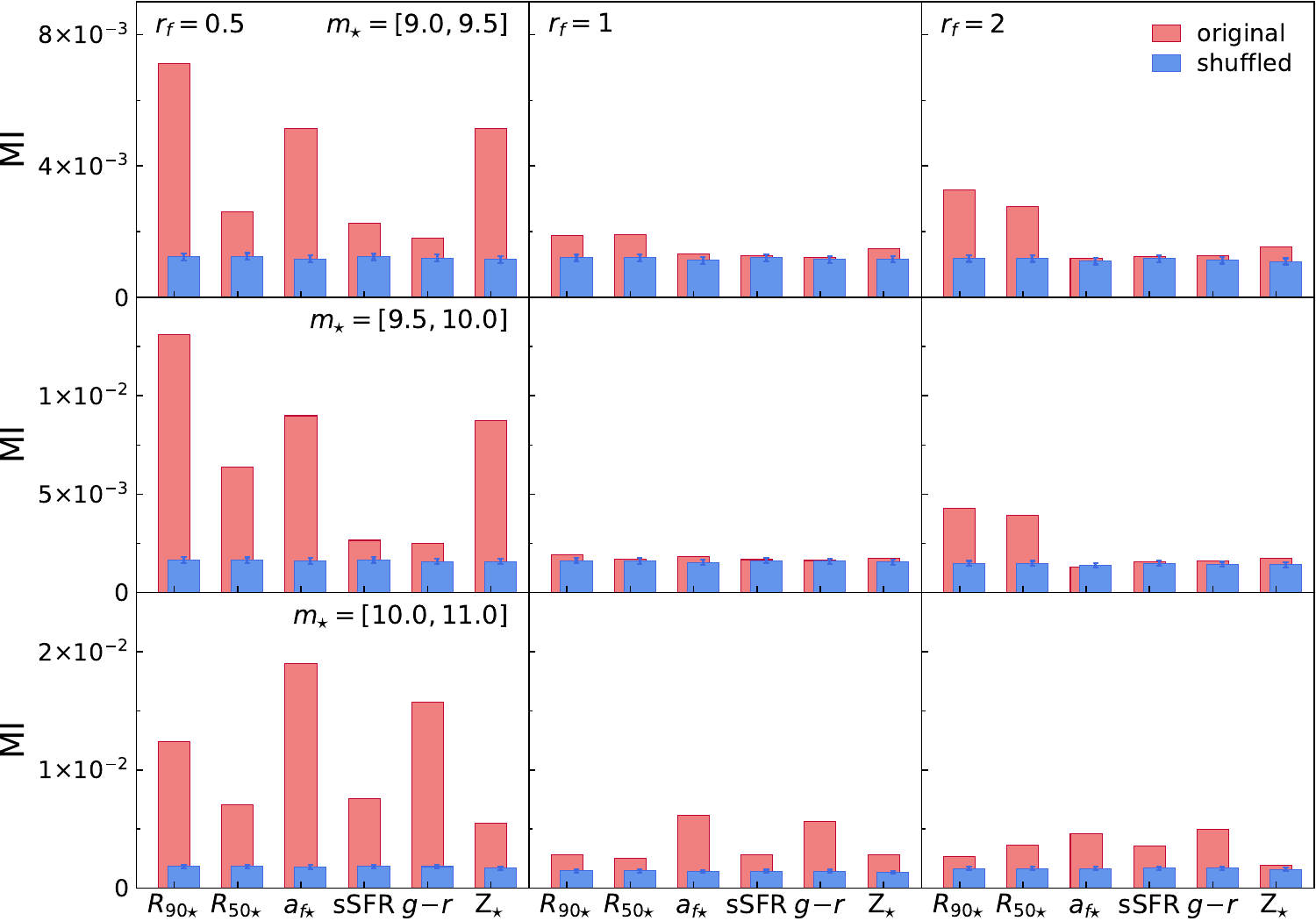}
% "\includegraphics" is very powerful; the graphicx package is already loaded
\caption{\label{fig:mcon} Same as figure~\ref{fig:ori} but from the $M_{\rm t}$-controlled sample where possible effects of mass differences are all reduced to 
a negligible level.}
\end{figure}
%%%%%%%%%%%%%%%%%%%%%%%%%%%%%%%%%%%%%%%%%%%%%%%%%%%%%%%%%%%%%%%%

%%%%%%%%%%%%%%%%%%%%%%%%%%%%%%%%%%%%%%%%%%%%%%%%%%%%%%%%%%%%%%%%%%%%%%%%%%%%

\subsection{Mutual information from the original galaxy sample}\label{sec:nocon}

To evaluate the mutual information (MI) between a galaxy stellar property (representatively say, $S$) and the primordial spin factor ($\tau$), we first divide 
two dimensional configuration space spanned by $S$ and $\tau$ into a total number of $N_{\rm pixel}$ pixels of small area. Let $N_{g}$ denote the total 
number of the selected target galaxies in the main sample. Let also $n(a,b)$ denote the number of those target galaxies whose values of $(S,\tau)$ belong to 
the $(a,b)$th pixel.  The ${\rm MI}(S,\tau)$, can be computed as
%%%%%%%%%%%%%%%%%%%%%%%%%%%%%%%%%%%%%%%%%%%%%%%%%%%%%%%%%%%%%%%%%%%%%%%%%%%%
\begin{equation}
\label{eqn:MI}
{\rm MI}(S,\tau) = \sum_{a=1}^{N_{a}}\sum_{b=1}^{N_{b}}P(S_{a},\tau_{b})\log \frac{P(S_{a},\tau_{b})}{P(S_{a})P(\tau_{b})}\, ,
\end{equation}
%%%%%%%%%%%%%%%%%%%%%%%%%%%%%%%%%%%%%%%%%%%%%%%%%%%%%%%%%%%%%%%%%%%%%%%%%%%%
Here, $(S_{a},\tau_{b})$ represent the median of $(S,\tau)$ values in the $(a,b)$th pixel, while $P(S_{a},\tau_{b})\equiv n(a,b)/N_{\rm g}$, 
$P(S_{a})\equiv \sum_{b=1}^{N_{b}}\,n(a,b)/N_{\rm g}$, and $P(\tau_{b})\equiv \sum_{a=1}^{N_{\rm a}}\,n(a,b)/N_{\rm g}$, where $N_{a}$ and $N_{b}$ 
denote the total numbers of bins into which the ranges of $S$ and $\tau$ values are divided, satisfying $N_{a}N_{b}=N_{\rm pixel}$.
A larger amount of ${\rm MI}(S,\tau)$ translates into the existence of a stronger connection between $S$ and $\tau$. 
To evaluate the statistical significance of MI values, we create $1000$ resamples of the target galaxies by randomly shuffling their $S$ and $\tau$ values, and 
compute the average and standard deviation over the $1000$ resamples for each property. 

Given the existence of strong correlations between $S$ and $M_{\star}$~\cite{gal-etal05,zhe-etal07,bot-etal16},  we classify the main sample into {\it low-}, {\it intermediate-}, 
and {\it high-mass} galaxies corresponding to three stellar mass ranges of $9\le m_{\star}< 9.5$, $9.5\le m_{\star}< 10$ and $10\le m_{\star}< 11$ with 
$m_{\star}\equiv M_{\star}/(h^{-1}M_{\odot})$, respectively, and separately compute the mean amounts of ${\rm MI}(S,\tau)$ by eq.~(\ref{eqn:MI}) in each $m_{\star}$-range.
Table~\ref{tab:mdq} lists the numbers ($N_{g}$) and mean Lagrangian sizes ($q_{50}$ and $q_{90}$) of the target galaxies falling in the three $m_{\star}$-ranges, where 
$q_{50}$ and $q_{90}$ are obtained from the spatial extents of the particles that occupy the $50\%$ and $90\%$ of $M_{t}$ from the center of mass in Lagrangian space, respectively. 
Figure~\ref{fig:ori} plots the amounts of MI (red histogram) from the original sample, and average MI over the shuffled resamples with one standard deviation errors 
(blue histogram) versus the six stellar properties in the three different ranges of $m_{\star}$ for the three different case of $r_{f}$.  As can be seen, for the case of $r_{f}=0.5$, 
all of the six stellar properties exhibit statistically significant signals of MI in all of the three stellar mass ranges except for the stellar metallicities that show negligibly low MI 
for the case of high-mass galaxies.  The amounts of MI increase as $m_{\star}$ increases and that the largest amount of MI is produced by $R_{90\star}$ in all of the three 
$m_{\star}$ ranges. The low and intermediate-mass galaxies exhibit an overall trend that the amounts and statistical significances of MI are lower for the case of larger $r_{f}$. 
Whereas, the high-mass galaxies yield the least significant amounts of MI on the scale of $r_{f}=1$ rather than on $r_{f}=2$, which implies that the initial density and tidal 
field defined on the larger scales ($r_{f}\ge 2$) also contribute to the establishments of the high-mass galaxy stellar properties.  

\subsection{Mutual information from controlled galaxy samples}\label{sec:con}
%%%%%%%%%%%%%%%%%%%%%%%%%%%%%%%%%%%%%%%%%%%%%%%%%%%%%%%%%%%%%%%%
\begin{table}[tbp]
\centering
\begin{tabular}{cccc}
\hline
\hline
\rule{0pt}{4ex} 
$m_{\star}$ & $N_{g}$ & $q_{50}$ & $q_{90}$ \medskip\\
 &  & $[h^{-1}{\rm Mpc}]$ & $[h^{-1}{\rm Mpc}]$ \medskip\\
\hline
\rule{0pt}{4ex}    
$[9.0, 9.5]$ & $54911$ & $0.74 \pm 0.11$ & $1.13 \pm 0.20$ \medskip\\
$[9.5, 10.0]$ & $41619$ & $0.89 \pm 0.13$ & $1.35 \pm 0.25$ \medskip\\
$[10.0, 11.0]$ & $48365$ & $1.33 \pm 0.36$ & $2.00 \pm 0.56$ \medskip\\
\hline
\end{tabular}
\caption{\label{tab:mdq} Total numbers and mean Lagrangian sizes of the target galaxies that fall in three different stellar mass ranges.}
\end{table}
%%%%%%%%%%%%%%%%%%%%%%%%%%%%%%%%%%%%%%%%%%%%%%%%%%%%%%%%%%%%%%%%

%%%%%%%%%%%%%%%%%%%%%%%%%%%%%%%%%%%%%%%%%%%%%%%%%%%%%%%%%%%%%%%%
\begin{table}[tbp]
\centering
\begin{tabular}{cccc}
\hline
\hline
\rule{0pt}{4ex} 
$m_{\star}$ & $N_{g}(r_{f}=0.5)$ & $N_{g}(r_{f}=1)$ & $N_{g}(r_{f}=2)$  \medskip\\
 &  & &  \medskip\\
\hline
\rule{0pt}{4ex}    
$[9.0, 9.5]$ & $23778$ & $25596$ & $24912$ \medskip\\
$[9.5, 10.0]$ & $16338$ & $16284$ & $17424$  \medskip\\
$[10.0, 11.0]$ &  $14322$ & $16854$ & $13326$  \medskip\\
\hline
\end{tabular}
\caption{\label{tab:mdqcon} Total number of the galaxies in the controlled samples for three different cases of $r_{f}$ in three different stellar mass ranges.}
\end{table}
%%%%%%%%%%%%%%%%%%%%%%%%%%%%%%%%%%%%%%%%%%%%%%%%%%%%%%%%%%%%%%%%

%%%%%%%%%%%%%%%%%%%%%%%%%%%%%%%%%%%%%%%%%%%%%%%%%%%%%%%%%%%%%%%%
\begin{figure}[tbp]
\centering % \begin{center}/\end{center} takes some additional vertical space
\includegraphics[width=0.85\textwidth=0 380 0 200]{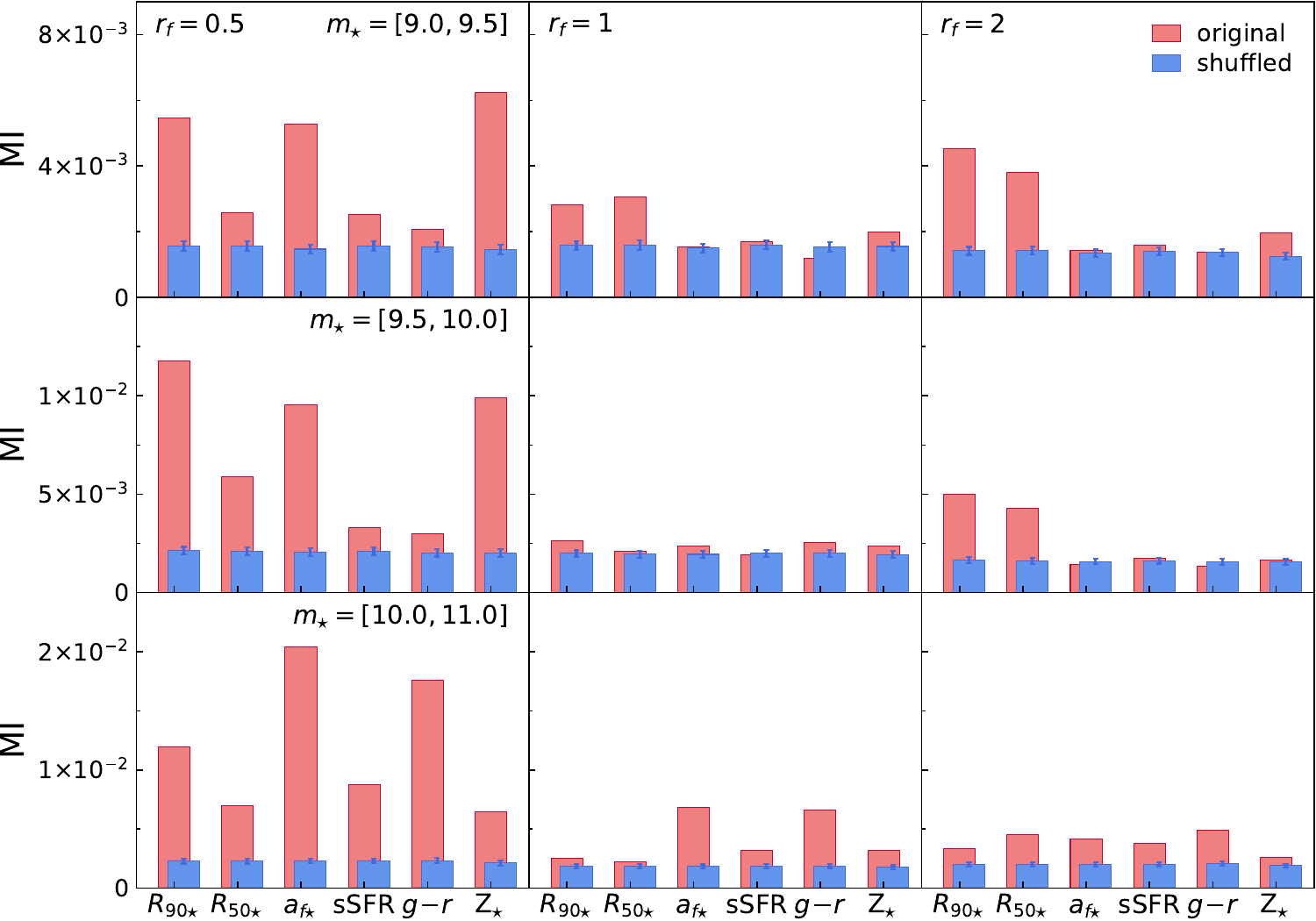}
% "\includegraphics" is very powerful; the graphicx package is already loaded
\caption{\label{fig:mdcon} Same as figure~\ref{fig:ori} but from the $M_{\rm t}$ and $\delta_{\rm nl}$ controlled sample in which any possible effects caused by 
differences in total masses and environmental densities are all reduced to a negligible level.}
\end{figure}
%%%%%%%%%%%%%%%%%%%%%%%%%%%%%%%%%%%%%%%%%%%%%%%%%%%%%%%%%%%%%%%%

The galaxy stellar properties are well known to be a function of the total mass, $M_{\rm t}$, of its host DM halo~\cite{bos-etal03,sha-etal06}. 
Furthermore, the DM halos with different total masses were found to have different $\tau$ values~\cite{ML24a},  which implies that the MI signals displayed in figure~\ref{fig:ori} 
could be at least partially contributed by the differences in $M_{\rm t}$ among the galaxies belonging to different $S$-$\tau$ pixels. To nullify a possible effect of mass difference 
on ${\rm MI}(S,\tau)$, it is necessary to eliminate $M_{\rm t}$ differences depending on the $\tau$ values. 
Binning the ranges of $\log m_{\rm t}$ and $\tau$, we count the number of galaxies whose values of $m_{\rm t}$ fall in each bin. 
Then, we look for a $\tau$ bin which yields the lowest galaxy number (say, $n_{\rm min}$) at a given $m_{\rm t}$ bin. 
Selecting only $n_{\rm min}(\log m_{\rm t})$ galaxies from each of the $\tau$ bins,  we end up having a controlled galaxy sample where the effect of mass differences 
among different pixels disappear.  Using this controlled sample, we recalculate the MI values by following the same procedure described in section~\ref{sec:nocon}. 

Figure~\ref{fig:mcon} shows the same as figure~\ref{fig:ori} but from the $m_{\rm t}$-controlled sample. As can be seen, although the controlled sample tends to yield slightly 
lower amounts of MI between the galaxy stellar properties and primordial spin factors, the signals are still quite significant for the case of $r_{f}=0.5$, confirming that 
$m_{\rm t}$-difference among the galaxies in the uncontrolled sample contribute at most only partially to the MI values. 
Note that the stellar ages and colors of the high-mass galaxies and the two stellar sizes of the low and intermediate-mass galaxies from the $m_{\rm t}$ controlled sample yield even 
larger and more significant amounts of MI for the cases of $r_{f}=0.5$ and $2$, respectively, than those from the original uncontrolled sample. This result implies that 
the $m_{\rm t}$ dependences of these stellar properties in fact played the role of veiling their connections with the primordial spin factors. 
%%%%%%%%%%%%%%%%%%%%%%%%%%%%%%%%%%%%%%%%%%%%%%%%%%%%%%%%%%%%%%%%
\begin{figure}[tbp]
\centering % \begin{center}/\end{center} takes some additional vertical space
\includegraphics[width=0.85\textwidth=0 380 0 200]{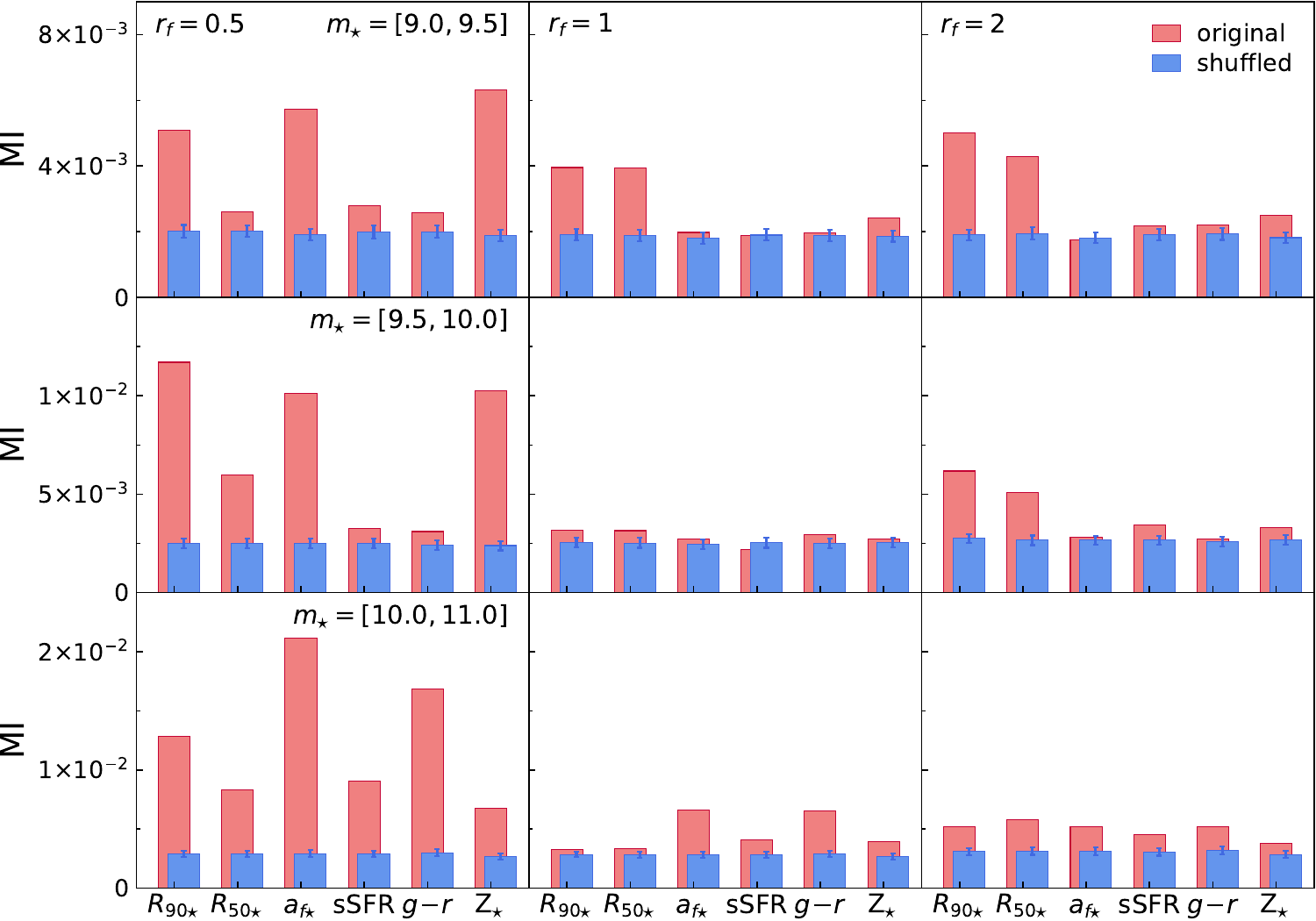}
% "\includegraphics" is very powerful; the graphicx package is already loaded
\caption{\label{fig:mdqcon} Same as figure~\ref{fig:ori} but from the $M_{\rm t}$, $\delta_{\rm nl}$ and $q$-controlled sample in which any possible 
effects caused by differences in total masses, environmental densities and shears are all reduced to a negligible level.}
\end{figure}
%%%%%%%%%%%%%%%%%%%%%%%%%%%%%%%%%%%%%%%%%%%%%%%%%%%%%%%%%%%%%%%%

Recalling that the galaxy stellar properties also exhibit strong variations with environmental density contrasts, $\delta_{\rm nl}$~\cite{gom-etal03,tan-etal04}, 
we further control the target galaxy sample to have identical joint distributions of $(m_{\rm t},\delta_{\rm nl})$, where the local density contrast field $\delta_{\rm nl}$ is 
reconstructed by applying the cloud-in-cell method to the TNG particle snapshot at $z=0$ in the same manner used for the reconstruction of the initial density field. 
Figure~\ref{fig:mdcon} shows the same as figure~\ref{fig:ori} but from the $m_{\rm t}$ and $\delta_{\rm nl}$-controlled sample. As can be seen, no drastic change 
is witnessed after the $\delta_{\rm nl}$ differences are eliminated. In all of the three $m_{\star}$ ranges, the amounts of MI between $R_{90\star}$ and $\tau$ are substantially 
reduced for the case of $r_{f}=0.5$, while for the low-mass galaxies, the statistical significances of MI between $R_{90\star}$ and $\tau$ is enhanced for the case of $r_{f}=2$. 

To be as scrupulous and conservative as possible in detecting a signal of net $\tau$-dependences of the galaxy stellar properties, we control the galaxy sample even more 
strictly to have identical joint distributions of $(m_{\rm t},\delta_{\rm nl}, q)$ among different $(S,\tau)$ pixels, where $q$ is the environmental shear~\cite{par-etal18} on which the 
galaxy properties were shown to depend~\cite{LL08,lee18,nan-etal24}, defined as~\cite{PS20}:
%%%%%%%%%%%%%%%%%%%%%%%%%%%%%%%%%%%%%%%%%%%%%%%%%%%%%%%%%%%%%%%%%%%%%%%%%%%%%%%%%%
\begin{equation}
\label{eqn:q}
q\equiv \bigg{\{}\frac{1}{2}\left[(\varrho_{1}-\varrho_{2})^{2} + (\varrho_{2}-\varrho_{3})^{2}+(\varrho_{3}-\varrho_{1})^{2}\right]\bigg{\}}^{1/2}\, ,
\end{equation}
%%%%%%%%%%%%%%%%%%%%%%%%%%%%%%%%%%%%%%%%%%%%%%%%%%%%%%%%%%%%%%%%%%%%%%%%%%%%%%%%%%
%%%%%%%%%%%%%%%%%%%%%%%%%%%%%%%%%%%%%%%%%%%%%%%%%%%%%%%%%%%%%%%%
\begin{figure}[tbp]
\centering % \begin{center}/\end{center} takes some additional vertical space
\includegraphics[width=0.85\textwidth=0 380 0 200]{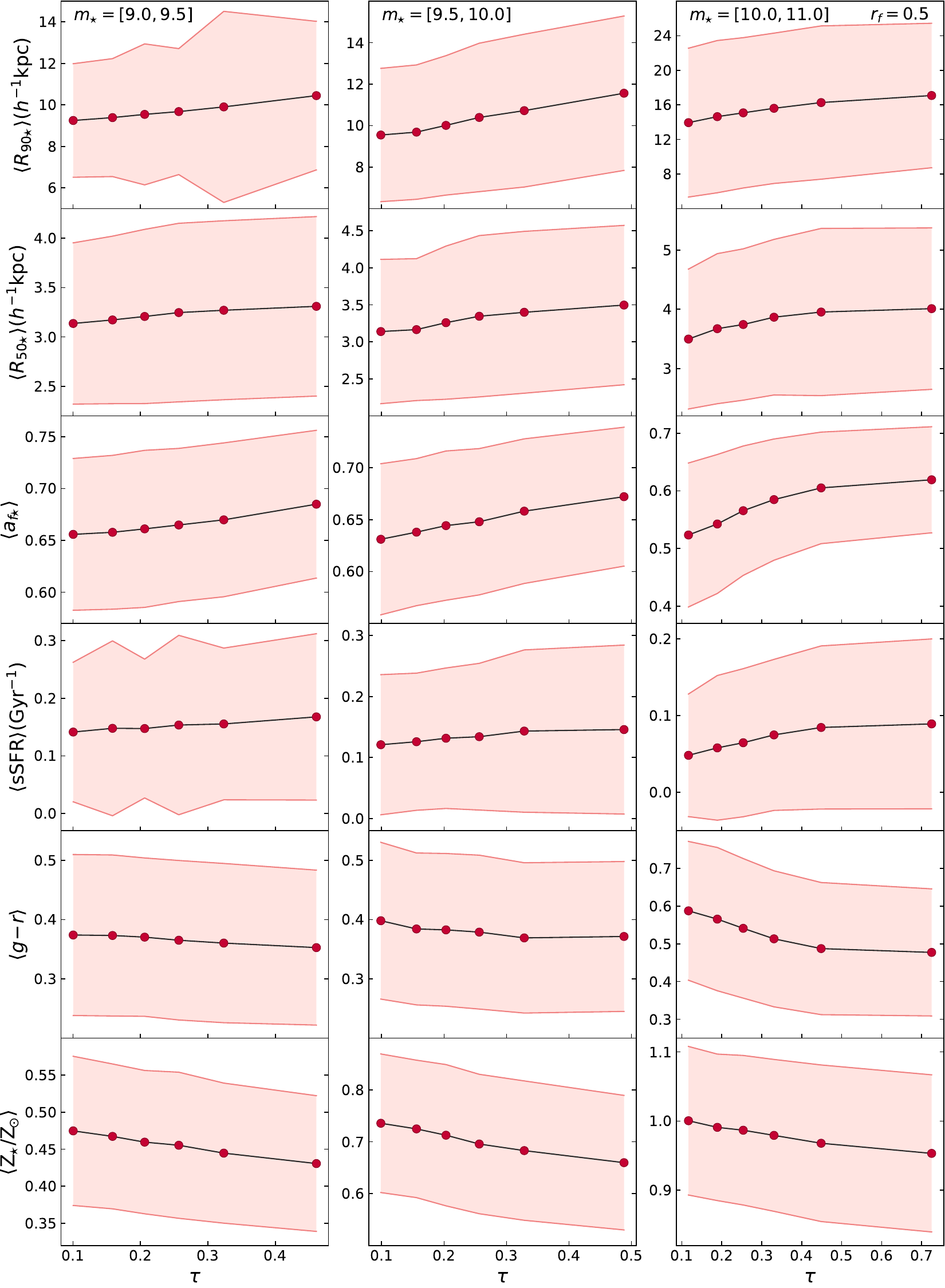}
% "\includegraphics" is very powerful; the graphicx package is already loaded
\caption{\label{fig:corr} Means (red filled circles) and $1\sigma$ standard scatter (pink area) of the six stellar properties averaged over the $M_{t}$ and $\delta$ controlled 
samples at six different $\tau$ bins for the case of $r_{f}=0.5$ in the three different $m_{\star}$ ranges.}
\end{figure}
%%%%%%%%%%%%%%%%%%%%%%%%%%%%%%%%%%%%%%%%%%%%%%%%%%%%%%%%%%%%%%%%
where $\{\varrho_{1},\ \varrho_{2},\ \varrho_{3}\}$ are three eigenvalues of the local tidal field measured at $z=0$. The local tidal field is reconstructed for the determination 
of $q$ from $\delta_{\rm nl}$ in the same manner used to construct $(T_{ij})$ from $\delta$, under the assumption that the magnitude of vorticity is negligible on the scale 
$r_{f}\ge 0.5$.  Table~\ref{tab:mdqcon} lists the numbers of the target galaxies included in the controlled samples for the 
three cases of $r_{f}$ in the three $m_{\star}$-ranges. As can be read, the sizes of the controlled samples depend on $r_{f}$, since 
the $\tau$ values for a galaxy varies with $r_{f}$.

Figure~\ref{fig:mdqcon} shows the same as figure~\ref{fig:ori} but from the $m_{\rm t}$, $\delta_{\rm nl}$ and $q$-controlled sample. As can be seen, 
the strictly controlled sample still yield significant signals of MI between the galaxy stellar properties and primordial spin factors, despite that possible effects of 
$M_{\rm t}$, $\delta_{\rm nl}$ and $q$ are all eliminated. It is interesting to notice that it varies with $m_{\star}$ which stellar property among the six has 
the largest amounts of ${\rm MI}(\tau)$ and how many stellar properties yield statistically significant MI signals. The overall trends are that for the case of $r_{f}=0.5$, 
the more stellar properties yield stronger signals, as $m_{\star}$ increases, while $R_{90\star}$ is $m_{\star}$-independent signals of MI. 
Another interesting aspect of the results shown in figure~\ref{fig:mdqcon} is that the significant signals of ${\rm MI}(\tau)$ seem to be contained in the two stellar sizes of the low 
and intermediate-mass galaxies, and in $g-r$ colors of the high-mass galaxies for the case of $r_{f}=1$ and $2$, respectively. 
The results shown in figure~\ref{fig:mdqcon} compellingly demonstrates that the stellar properties of present galaxies have significant net $\tau$ dependences, 
independent of total mass and environments, whose establishments should be ascribed to the multi-scale influences of initial density and tidal fields at the protogalactic 
stages. 

Now that the galaxy stellar properties are found to share significant amounts of ${\rm MI}(\tau)$, it should also be worth investigating whether each of $S$ is positively or negatively 
correlated with $\tau$. For this investigation, we split the $\tau$-values into $six$ short intervals, and then take the average and $1\sigma$ scatter of each $S$  
over the galaxies in each $\tau$ bin.  Figure~\ref{fig:corr} shows the results for the case of $r_{f}=0.5$ in the three ranges of $m_{\star}$, using 
the $M_{t}$ and $\delta$ controlled galaxy samples. 
As can be seen, the four of the six properties ($R_{90\star}$, $R_{50\star}$, $a_{f\star}$, sSFR) are positively correlated with $\tau$ while the other two ($g-r$ and $Z_{\star}$) are 
negatively correlated with $\tau$, regardless of $m_{\star}$. In other words, the protogalaxies formed at the initial sites with larger $\tau$ values tend to evolve into the galaxies 
with larger sizes, higher specific star formation rates, forming later, bluer colors and lower stellar metallicities. 

\subsection{A bimodal $\Gamma$-distribution of galaxy stellar sizes}\label{sec:rdist}
%%%%%%%%%%%%%%%%%%%%%%%%%%%%%%%%%%%%%%%%%%%%%%%%%%%%%%%%%%%%%%%%
\begin{figure}[tbp]
\centering % \begin{center}/\end{center} takes some additional vertical space
\includegraphics[width=0.85\textwidth=0 380 0 200]{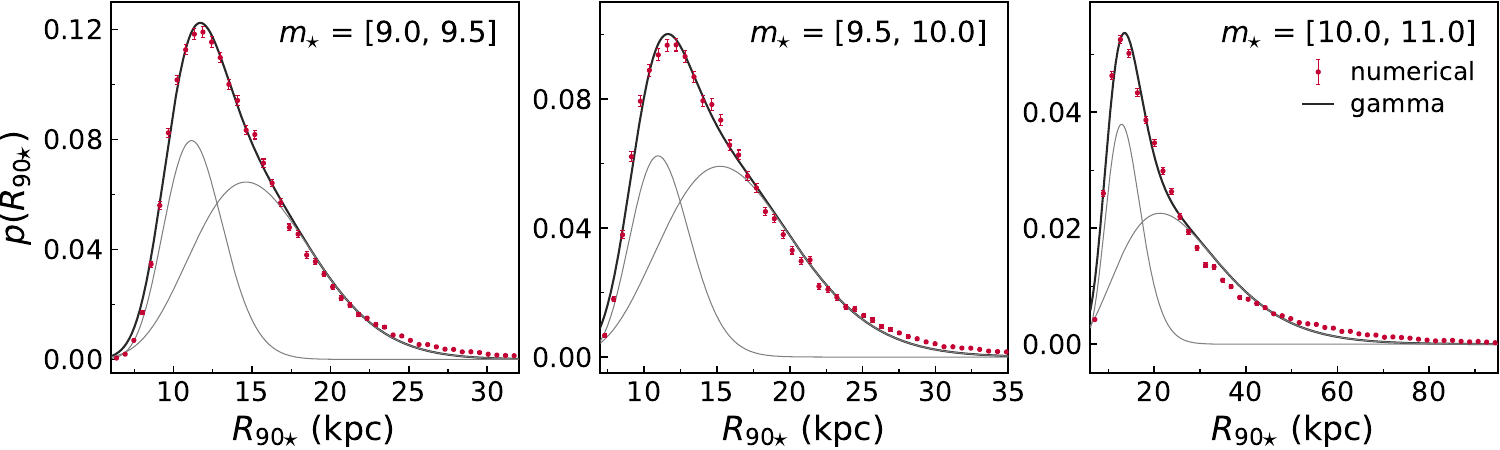}
% "\includegraphics" is very powerful; the graphicx package is already loaded
\caption{\label{fig:gam} Probability density function of the galaxy stellar sizes with bootstrap errors (filled circles) compared with 
a bimodal model (thick solid lines) consisting of two different $\Gamma$ distributions (thin solid lines) in each of three stellar mass ranges.}
\end{figure}
%%%%%%%%%%%%%%%%%%%%%%%%%%%%%%%%%%%%%%%%%%%%%%%%%%%%%%%%%%%%%%%%

%%%%%%%%%%%%%%%%%%%%%%%%%%%%%%%%%%%%%%%%%%%%%%%%%%%%%%%%%%%%%%%%
\begin{table}[tbp]
\centering
\begin{tabular}{cccc}
\hline
\hline
\rule{0pt}{4ex} 
$m_{\star}$ & ($k_{1},\ \theta_{1})$ & $\xi$ & ($k_{2},\ \theta_{2})$ \medskip \\ 
\hline
\rule{0pt}{4ex}    
$[9.0, 9.5]$ & $(15.57\pm 5.11,\ 1.00\pm 0.05)$ & $0.38\pm 0.09$ & $(36.14\pm 2.39,\ 0.32\pm 0.12)$ \medskip \\
$[9.5, 10.0]$ & $(12.10\pm 2.14,\ 1.37\pm 0.19)$ & $0.32\pm 0.10$ & $(30.17\pm 5.70,\ 0.38\pm 0.07)$ \medskip \\
$[10.0, 11.0]$ & $(4.70\pm 2.79,\ 5.75\pm 0.25)$ & $0.36\pm 0.11$ & $(12.86\pm 1.20,\ 1.09\pm 1.04)$ \medskip \\
\hline
\end{tabular}
\caption{\label{tab:gam} Two sets of two best-fitparameters of two $\Gamma$ modes of the bimodal model for $p(R_{90\star})$ in three stellar mass ranges.}
\end{table}
%%%%%%%%%%%%%%%%%%%%%%%%%%%%%%%%%%%%%%%%%%%%%%%%%%%%%%%%%%%%%%%%

Given that the galaxy stellar sizes are strongly correlated with the halo spin parameter, $\lambda$~\cite{KL13}, and that the probability density function of $\lambda$ was 
found to be approximated much better by a $\Gamma$-distribution with two characteristic parameters~\cite{ML24a} rather than by the conventional log-normal 
model~\cite{bul-etal01}, we speculate that $R_{90\star}$ may also be well modeled by a $\Gamma$ distribution.
To verify this speculation, we determine the distribution of $R_{90\star}$ by taking the following steps.  Divide the range of $R_{90\star}$ into 
multiple differential intervals of equal size, $\Delta_{90\star}$. Count the numbers, $n_{\rm g}$,  of the galaxies whose values of $R_{90\star}$ fall in each interval, 
and then determine the probability density, $p(R_{90\star})$, at each differential interval as $n_{\rm g}/(N_{\rm g}\Delta_{90\star})$. 

Figure~\ref{fig:gam} plots $p(R_{90\star})$ for the three cases of $m_{\star}$ range, revealing that $p(R_{90\star})$ has a long tail in the large size section 
($R_{90\star}> R_{90\star, c}$), but drops rapidly in the opposite section ($R_{90\star}\le R_{90\star, c}$) with size threshold in the range of 
$15\le R_{90\star,c}/{\rm kpc}\le 20$.
Comparing this numerically determined $p(R_{90\star})$ to a $\Gamma$ distribution, we find that unlike the case of $p(\lambda)$, a single $\Gamma$
distribution fails to match simultaneously the stellar size distribution $p(R_{90\star})$ in the whole range of $R_{90\star}$. 
Instead, a separate treatment of two ranges,  $R_{90\star}\le R_{90\star, c}$ and $R_{90\star}> R_{90\star, c}$, in fitting $p(R_{90\star})$ to a single $\Gamma$ 
distribution turns out to work well, yielding two different sets of characteristic parameters.  In other words, the stellar size distribution, $p(R_{90\star})$, turns out to 
have two different $\Gamma$ modes, and this fitting result leads us to discover that the following bimodal Gamma distribution describes quite well $p(R_{90\star})$ 
in the whole range of $R_{90\star}$:
%%%%%%%%%%%%%%%%%%%%%%%%%%%%%%%%%%%%%%%%%%%%%%%%%%%%%%%%%%%%%%%%%%%%%%%%%%%%%%%%%%%%
\begin{eqnarray}
\label{eqn:bim}
p(R_{90\star}) &=& \xi\times \Gamma\left(R_{90\star}; k_{1},\theta_{1}\right) + (1-\xi)\times \Gamma\left(R_{90\star}; k_{2},\theta_{2}\right)\, , \\
\label{eqn:fra}
\Gamma\left(R_{90\star}; k_{i},\theta_{i}\right) &=& \frac{1}{f(k_{i})\theta^{k_{i}}_{i}}R_{90\star}^{k_{i}-1}\exp\left(-\frac{R_{90\star}}{\theta_{i}}\right)\, ,\,\, {\rm for}\,\, i\in\{1,2\} \\
\label{eqn:gam}
f(k_{i}) &=& \int_0^{\infty} t^{k_{i}-1}e^{-t}\,dt\, .
\end{eqnarray}
%%%%%%%%%%%%%%%%%%%%%%%%%%%%%%%%%%%%%%%%%%%%%%%%%%%%%%%%%%%%%%%%%%%%%%%%%%%%%%%%%%%%
Here, $\xi$ and $1-\xi$ represent the fractions of the first and second $\Gamma$ distributions with characteristic parameters $(k_{1},\theta_{1})$ and $(k_{2},\theta_{2})$, 
respectively.  

Employing the $\chi^{2}$-statistics\footnote{Technically, the python package, ${\rm ``scipy.optimize.curve\_fit"}$ is utilized for the computation of the best-fit parameters and 
associated errors for the two $\Gamma$ distributions.}, we fit the numerically obtained $p(R_{90\star})$ to eqs.~(\ref{eqn:bim})-(\ref{eqn:gam}) by adjusting the values 
of $(k_{1},\ \theta_{1})$ and $(k_{2},\ \theta_{2})$ as well as $\xi$. Figure~\ref{fig:gam} compares the best-fit bimodal $\Gamma$ distributions (thick solid lines) with the 
numerical results (red filled circles), confirming its validity in all of the three $m_{\star}$-ranges. Table~\ref{tab:gam} lists the best-fit parameter values for the three cases of 
$m_{\star}$ ranges. 
To physically understand this bimodal nature of $p(R_{90\star})$, recall the previous result that the probability density function of $\tau$ was also found to follow a single 
$\Gamma$-distribution whose best-fit parameters $(k,\ \theta)$ turned out to be scale dependent~\cite{ML24b}.  In other words, if $\tau$ is measured from the initial tidal fields 
smoothed on a different scale, then $p(\tau)$ is described by a single $\Gamma$ distribution with a different set of two parameters.  Given this previous finding,  we interpret the 
bimodal nature of $p(R_{90\star})$ as an evidence supporting that the galaxy stellar sizes $R_{90\star}$ are affected by multi-scale initial conditions. 

\section{MI between the protogalaxy angular momenta and galaxy stellar properties}\label{sec:proto}
%%%%%%%%%%%%%%%%%%%%%%%%%%%%%%%%%%%%%%%%%%%%%%%%%%%%%%%%%%%%%%%%
\begin{figure}[tbp]
\centering % \begin{center}/\end{center} takes some additional vertical space
\includegraphics[width=0.85\textwidth=0 380 0 200]{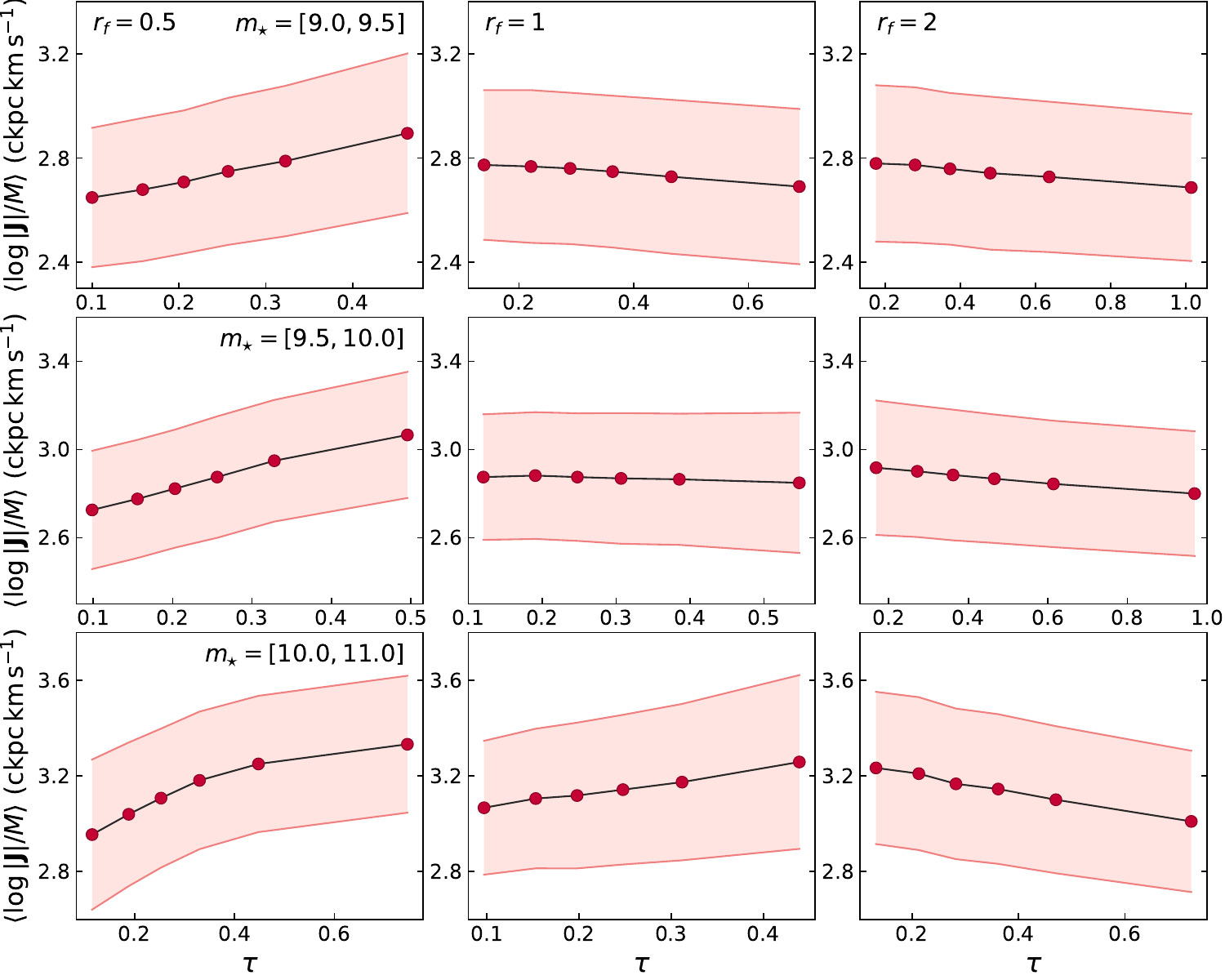}
% "\includegraphics" is very powerful; the graphicx package is already loaded
\caption{\label{fig:proto_cor} Means (red filled circles) and $1\sigma$ standard scatter (pink area) of the specific angular momenta of protogalactic sites at six different 
$\tau$ bins for three different cases of $r_{f}$ in three different $m_{\star}$ ranges.}
\end{figure}
%%%%%%%%%%%%%%%%%%%%%%%%%%%%%%%%%%%%%%%%%%%%%%%%%%%%%%%%%%%%%%%%
%%%%%%%%%%%%%%%%%%%%%%%%%%%%%%%%%%%%%%%%%%%%%%%%%%%%%%%%%%%%%%%%
\begin{figure}[tbp]
\centering % \begin{center}/\end{center} takes some additional vertical space
\includegraphics[width=0.85\textwidth=0 380 0 200]{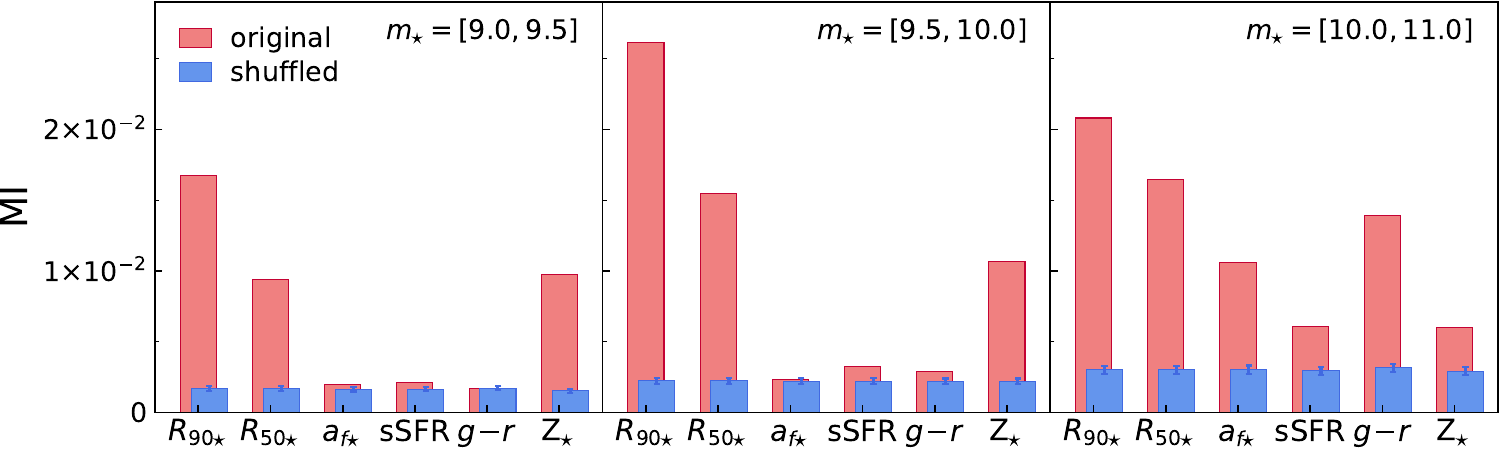}
% "\includegraphics" is very powerful; the graphicx package is already loaded
\caption{\label{fig:proto_MI} MI values with $1\sigma$ scatters between the six stellar properties of the TNG central galaxies at $z=0$ and the protogalaxy 
specific angular momenta, $\vert{\bf J}\vert/M_{t}$, measured at $z_{i}=127$ from the $M_{\rm t}$ and $\delta_{\rm nl}$ controlled samples in three different $m_{\star}$ ranges.}
\end{figure}
%%%%%%%%%%%%%%%%%%%%%%%%%%%%%%%%%%%%%%%%%%%%%%%%%%%%%%%%%%%%%%%%

In the framework of the linear tidal torque theory~\cite{dor70,whi84}, it is natural to expect that the significant amounts of ${\rm MI}(\tau)$ shared by the six galaxy stellar 
properties may be caused by the correlations between $\tau$ and magnitudes of protogalaxy angular momenta and thus that the six galaxy stellar properties must share very similar 
mutual information with the magnitude of protogalaxy angular momenta. 
To verify this expectation, we compute the correlations between $\tau$ and magnitudes of the protogalaxy specific angular momenta, $\vert{\bf J}\vert/M$, in each of the three 
$m_{\star}$-ranges in the same way as in section~\ref{sec:con}, the results of which are shown in figure~\ref{fig:proto_cor}. 
As can be seen,  for the case of $r_f=0.5$, the protogalaxy angular momenta indeed exhibit positive correlations with $\tau$ regardless of $m_{\star}$. 
In the stellar mass range of $m_{\star}\le 10$ is found an almost linear positive correlation, while a deviation from the linear correlation trend is found around $\tau=0.5$ in the higher 
$m_{\star}$-range. For the case of $r_{f}=1$, a positive linear correlation is found in the range of $10\le m_{\star}\le 11$, while a negative linear correlation and no correlation are  
found from the galaxies with $m_{\star}\le 9.5$ and with $9.5\le m_{\star}<10$, respectively.  For the case of $r_{f}=2$, all of the stellar mass ranges yield linear negative correlations 
between $\tau$ and $\vert{\bf J}\vert/M$, the strengths of which appear to increase with $m_{\star}$.

We also evaluate the mutual informations between the six galaxy stellar properties and $\vert{\bf J}\vert/M$ by the same procedure as described in section~\ref{sec:con}, 
the results of which are shown in figure~\ref{fig:proto_MI}.  In the range of $m_{\star}\le 10$, a significant signal of ${\rm MI}(\vert{\bf J}\vert/M)$ is yielded only by 
$R_{90\star}$, $R_{50\star}$ and $Z_{\star}$, while in the range of $10< m_{\star}\le 11$, all of the six properties are found to share the significant amounts of 
${\rm MI}(\vert{\bf J}\vert/M)$.  The comparison of ${\rm MI}(\tau)$ shown in figure~\ref{fig:mdcon} with this result, ${\rm MI}(\vert{\bf J}\vert/M)$ reveals appreciably different 
behaviors between the two.  In particular, the formation epochs of the galaxies with $m_{\star}\le 10$ yield significant amounts of mutual information with $\tau$ and ${\rm MI}$ 
for the case of $r_{f}=0.5$, but negligibly little amount of mutual information with $\vert{\bf J}\vert/M$.

A crucial implication of the results shown in figures~\ref{fig:proto_cor}--\ref{fig:proto_MI} is that the significant amounts of ${\rm MI}(\tau)$ shared by the six stellar properties are 
not solely caused by the effects of protogalaxy angular momenta on the galaxy stellar properties. In other words, the roles of $\tau$ in establishing the galaxy stellar properties 
are not limited to promoting the protogalaxy angular momenta but extended to provoking and regulating multi-scale tidal effects on the protogalaxies, which strongly affects the formation 
epochs of descendant galaxies. Besides, unlike the protogalaxy angular momenta, the $\tau$ values of real galaxies are in principle estimable from observations, as mentioned in 
section~\ref{sec:intro}.  Henceforth, we suggest that the primordial spin factor should be a more appropriate and practical single parameter to quantify the extensive 
effects of initial conditions on the galaxy stellar properties beyond promoting the protogalaxy angular momenta.  

\section{Summary and conclusion}\label{sec:sum}

In light of the recent numerical results that the spin parameters and formation epochs of galactic subhalos exhibited strong dependences on the single parameter initial condition for 
the first order generation of protogalaxy angular momentum~\cite{ML24a}, we have numerically investigated if and how strongly the single parameter initial condition also 
contributes to the establishment of six observable galaxy properties: stellar sizes ($R_{90\star}$ and $R_{50\star}$), stellar ages ($a_{f\star}$), specific star formation rates (sSFR), 
optical $g-r$ colors ($g-r$), and metallicities ($Z_{\star}$). As in our prior work~\cite{ML24a}, the current analysis has been focused on the sample of central galaxies in the stellar 
mass range of  $9\le m_{\star}\equiv \log\left[M_{\star}/(h^{-1}M_{\odot})\right]\le 11$ at $z=0$ from the TNG300-1 hydrodynamic simulations, adopting the tidal torque picture 
according to which the single parameter initial condition is quantified by the degree of misalignments between the initial tidal field $(T_{ij})$ and protogalaxy inertia tensors 
$(I_{ij})$~\cite{dor70,whi84,LP00}. 

Unlike in ref.~\cite{ML24a}, however, we have replaced $(I_{ij})$ by the density hessian tensor, $(H_{ij})$,  defined as the second derivative of the initial density field 
on two grounds:  $(H_{ij})$ was found to approximate $(I_{ij})$ quite well on the scale $R_{f}$ comparable to a protogalaxy Lagrangian radius~\cite{yu-etal20}, 
and the former is much more robust than the latter against the variation of subhalo finding algorithm and definition of subhalo virial radius. 
In addition, $(H_{ij})$ is observationally reconstructable at initial redshifts unlike the protogalaxy inertia tensors~\cite{mot-etal21}. 
Defining the single parameter initial condition as the degree of misalignment between $(H_{ij})$ and $(T_{ij})$, and calling it the primordial spin factor, $\tau$, we have 
measured the values of $\tau$ at the protogalactic sites of the TNG central galaxies at $z=127$. Given that the $\tau$ values vary with the smoothing scale, 
$r_{f}\equiv R_{f}/(h^{-1}{\rm Mpc})$, we have considered three different cases of $r_{f}=0.5,\ 1$ and $2$. 

The net $\tau$-dependences of the aforementioned six stellar properties have been evaluated by the amounts of their mutual information (MI), for which the controlled 
sample of the central galaxies have been created, where the effects of mass and environment differences are all minimized. It has been found that the amounts and 
statistical significances of the ${\rm MI}(\tau)$ signals from the six stellar properties as well as  the stellar property containing the largest amount of MI depend on 
$m_{\star}$ and $r_{f}$.  We have also determined the mean values of the six stellar properties as a function of $\tau$ to see whether they decrease or increase with $\tau$, 
and compare the $\tau$-dependences of the six stellar properties with their dependences on the magnitudes of protogalaxy angular momenta, $\vert{\bf J}\vert/M$ which are 
directly proportional to $\tau$ in the linear tidal torque theory~\cite{dor70,whi84}.  

In the following, we provide a summary of the key results of our analysis. 
%%%%%%%%%%%%%%%%%%%%%%%%%%%%%%%%%%%%%%%%%%%%%%%%%%%%%%%%%%%%%%%%%%%%%%%%%%%%%%%%%%%%%%
\begin{itemize}
\item
In the stellar mass range of $9\le m_{\star}< 9.5$,  only the three stellar properties, $R_{90\star}$, $a_{f\star}$ and $Z_{\star}$, among the six are found to contain 
significant amounts of ${\rm MI}(\tau)$ for the case of $r_{f}=0.5$.  The largest amount of MI about the primordial spin factor $\tau$ is exhibited by the stellar metallicity, 
$Z_{\star}$. Although the amount of MI between each of these three properties and $\tau$ shows a trend to diminish as $r_{f}$ increases, the $90\%$ stellar mass radius, 
$R_{90\star}$, still exhibits substantial amounts of ${\rm MI}(\tau)$ even at $r_{f}=1$ and $2$. 
\item
In the stellar mass range of $9.5\le m_{\star}< 10$,  the 50 percent stellar mass radius, $R_{50\star}$, as well as $R_{90\star}$, $a_{f\star}$ and $Z_{\star}$, 
yield significant amounts of ${\rm MI}(\tau)$ for the case of $r_{f}=0.5$.  Among these four, it is $R_{90\star}$ which shares the largest amount of MI with $\tau$. 
The probability density function, $p(R_{90\star})$, is shown to be well described by a bimodal $\Gamma$ distribution, which is interpreted as an an evidence of multi-scale 
influences of the initial conditions on the establishment of galaxy stellar properties. 
\item
In the stellar mass range of $10\le m_{\star}\le 11$,  significant MI signals are yielded for the case of $r_{f}=0.5$ by all of the six stellar properties, among which 
$a_{f\star}$, and $g-r$ colors yield the most significant amounts of ${\rm MI}(\tau)$.  
\item
The higher stellar mass a galaxy has, the larger amount of MI its stellar property shares with $\tau$. The only exception is the stellar metallicity, $Z_{\star}$, which 
yields a larger amount of ${\rm MI}(\tau)$ in the lower stellar mass range of $m_{\star}\le 10$ than in the higher counterpart.
\item
The amounts of ${\rm MI}(\tau)$ are found to decrease as $r_{f}$ increases for all of the stellar properties except for the 50 percent-mass radius, $R_{50\star}$, which exhibits a 
larger amount of ${\rm MI}(\tau)$ at the larger smoothing scales in the lowest stellar mass range $9\le m_{\star}\le 9.5$.
\item
Among the six stellar properties, the positive correlations with $\tau$ are exhibited by $R_{90\star}$, $R_{50\star}$, $a_{f\star}$ and sSFR, while negative correlations 
by the other two, $g-r$ color and $Z_{\star}$. 
\item
The existence of a strong correlation between $\tau$ and $\vert{\bf J}\vert/M$ seems to be responsible at least partially for the large amounts of ${\rm MI}(\tau)$ yielded by 
$R_{90\star}$, $R_{50\star}$ and $Z_{\star}$. But, it fails to account for the highly significant ${\rm MI}(\tau)$ yielded by $a_{f\star}$, regardless of $r_f$. 
\end{itemize}
%%%%%%%%%%%%%%%%%%%%%%%%%%%%%%%%%%%%%%%%%%%%%%%%%%%%%%%%%%%%%%%%%%%%%%%%%%%%%%%%%%%%%%

Before drawing a conclusion about the $\tau$-effects on the galaxy stellar properties,  however, it should be worth mentioning a caveat that concerns the validity of our results. 
The amount of ${\rm MI}(\tau)$ yielded by the galaxy stellar properties may well be subject to the issue of which baryon physics are included in hydrodynamic simulations considered. 
Despite that the TNG simulations comprehensively included all of the known physical processes of baryons from radiative cooling and heating through supernovae feedbacks to 
magnetic fields, the list should still be far from being complete~\cite{tng1,tng2,tng3,tng4,tng5}. 
If ${\rm MI}(\tau)$ were determined from a different hydrodynamical simulation which adopts a model of stronger baryon feedback, 
then its amount could diminish to a less significant level, as stronger feedback effects on the galaxies in the nonlinear regime are likely to reduce more severely 
the dependences of galaxy stellar properties on the initial conditions.  To properly address this issue,  what is required is an observational estimation of ${\rm MI}(\tau)$ 
from the reconstructable initial density and potential fields~\cite{mot-etal22}. If the observed signal turned out to be consistent with our results from the TNG simulations, 
then it would confirm the $\tau$-dependence of galaxy properties and help us constrain galaxy formation models, providing a whole new insight into the long-standing debate 
on the roles of nature vs. nurture for the establishments of present galaxy properties. 
We intend to pursue a follow-up work in this direction and hope to report the result in the near future.

\acknowledgments

The IllustrisTNG simulations were undertaken with compute time awarded by the Gauss Centre for Supercomputing (GCS) 
under GCS Large-Scale Projects GCS-ILLU and GCS-DWAR on the GCS share of the supercomputer Hazel Hen at the High 
Performance Computing Center Stuttgart (HLRS), as well as on the machines of the Max Planck Computing and Data Facility 
(MPCDF) in Garching, Germany.  JSM acknowledges the support by the National Research Foundation (NRF) of Korea grant 
funded by the Korean government (MEST) (No. 2019R1A6A1A10073437). JL acknowledges the support by Basic Science 
Research Program through the NRF of Korea funded by the Ministry of Education (No.2019R1A2C1083855). 

% The bibliography will probably be heavily edited during typesetting.
% We'll parse it and, using the arxiv number or the journal data, will
% query inspire, trying to verify the data (this will probalby spot
% eventual typos) and retrive the document DOI and eventual errata.
% We however suggest to always provide author, title and journal data:
% in short all the informations that clearly identify a document.

\end{document}